\documentclass[aps,superscriptaddress,reprint,twocolumn]{revtex4-1}
\usepackage[colorlinks, citecolor=blue, linkcolor=blue]{hyperref}
\usepackage[english]{babel}
\usepackage{amsmath}
\usepackage{amssymb}
\usepackage{mathtools}
\usepackage{array}
\usepackage{enumitem}
\usepackage{nicefrac}
\usepackage[separate-uncertainty = true]{siunitx}
\usepackage{graphicx}
\usepackage{xcolor}
\usepackage{xspace}

\usepackage{changes}
\newlist{todolist}{itemize}{2}
\setlist[todolist]{label=$\square$}
\usepackage{pifont}
\newcommand{\sub}[1]{\textsubscript{#1}}
\newcommand{\BFO}{BiFeO$_3$\xspace}

\begin{document}

\title{Spin-current driven Dzyaloshinskii-Moriya interaction in the multiferroic \BFO from first-principles}

\author{Sebastian Meyer}
\affiliation{Nanomat/Q-mat/CESAM, Universit\'e de Li\`ege, B-4000 Sart Tilman, Belgium}

\author{Bin Xu}
\affiliation{Jiangsu Key Laboratory of Thin Films, School of Physical Science and Technology, Soochow University, Suzhou 215006, China}
\affiliation{Physics Department and Institute for Nanoscience and Engineering, University of Arkansas, Fayetteville, Arkansas 72701, USA}

\author{Matthieu J. Verstraete}
\affiliation{Nanomat/Q-mat/CESAM, Universit\'e de Li\`ege, B-4000 Sart Tilman, Belgium}

\author{Laurent Bellaiche}
\affiliation{Physics Department and Institute for Nanoscience and Engineering, University of Arkansas, Fayetteville, Arkansas 72701, USA}

\author{Bertrand Dup\'e}
\affiliation{Nanomat/Q-mat/CESAM, Universit\'e de Li\`ege, B-4000 Sart Tilman, Belgium}
\affiliation{Fonds de la Recherche Scientifique (FRS-FNRS), Bruxelles, Belgium}

\date{Mars 2022}

\begin{abstract}

The electrical control of magnons opens up new ways to transport and process information for logic devices. In magnetoelectrical multiferroics, the Dzyaloshinskii-Moriya (DM) interaction directly allow for such a control and, hence, is of major importance.
We determine the origin and the strength of the (converse) spin current DM interaction in the \textit{R3c} bulk phase of the multiferroic \BFO based on density functional theory. Our data supports only the existence of one DM interaction contribution originating from the spin current model. By exploring thenmagnon dispersion in the full Brillouin Zone, we show that the exchange is isotropic, but the DM interaction and anisotropy prefer any propagation and any magnetization direction within the full (111) plane. Our work emphasizes the significance of the asymmetric potential induced by the spin current over the structural asymmetry induced by the anionic octahedron in multiferroics such as \BFO.

\end{abstract}

\maketitle

\paragraph*{Introduction} \BFO (BFO) is one of the few single-phase multiferroics which exhibit a large spontaneous polarization and a long range magnetic order at room-temperature. BFO has an antiferromagnetic (AFM) texture that can be approximated locally by a G-type order in its \textit{R3c} ground state with a N\'eel temperature of 643~K \cite{Roginskaya_Y_E_TomashpolSkii1966-lr}. \textit{R3c} BFO exhibits a polarization of about 90~$\mu$C/cm$^2$ along the pseudo-cubic $<111>$ symmetry equivalent directions with a Curie temperature of 1123~K \cite{Neaton2005-ti,Kornev2007-xg,Haumont2008-co,Ravindran2006-tu}.

In bulk, the AFM order is modified by the Dzyaloshinskii-Moryia (DM) interaction \cite{Dzialoshinskii1958-xd,Moriya1960-oy} creating a spin spiral that propagates along the $[1\bar{1}0]$ direction (known as type-I cycloid), with magnetic moments lying in the plane formed by the polarization and the propagation direction. The periodicity of this spin spiral is 62~nm \cite{Sosnowska1982,Sosnowska1996}. Additionally, another propagation direction along $[11\bar{2}]$ has been reported in BFO, referred to as type-II cycloid \cite{Ratcliff2011-rr,Sando2013,Bertinshaw2016-ri,Burns2019,Haykal2020}.

The AFM spin spiral couples to the polarization via the magnetoelectric (ME) effect. The ME effect can have multiple origins \cite{Martin2008-kg}. The DM interaction directly couples atomic displacements to the AFM spin spiral and opens up magnons electrical control \cite{Rovillain2010}. This effect is at the center of a new logic device which intents to electrically control out-of-equilibrium spin spirals - also called magnons - in BFO to transport and process information \cite{Manipatruni2019-pw,Parsonnet2022-op}. The exploration of the stability mechanisms of different types of spin spirals is of paramount importance and has been subjected to a lot of research.

A phenomenological model based on Lifschitz invariant established that the DM interaction stabilizing the spin spiral had the form
$\alpha ( P ) \cdot \mathbf{L}_i \times \mathbf{L}_j$ where $P$ is the polarization and $L$ is the AFM vector \cite{Sosnowska1995-xg}. This phenomenological model was completed by a microscopic model based on tight binding approximation which attributed the presence of a non-zero polarization to the presence of a spin spiral.
The hybridization between the \emph{d}-orbitals of the metal ions and the \emph{p}-orbitals of the Oxygen results in the polarization $\mathbf{P} \propto \mathbf{e}_{ij} \times ( \mathbf{S}_i \times \mathbf{S}_j ) $ where $\mathbf{S}_i$ and $\mathbf{S}_j$ are spins on site $i$ and $j$, respectively and $\mathbf{e}_{ij}$ is the unit vector between site $i$ and $j$ \cite{Katsura2005}. The link between this polarization and the presence of the DM interaction was explicitly written by Rahmedov \emph{et al.} where the spin spiral in BFO was explained based on the presence of two chiral interactions of different symmetries \cite{Rahmedov2012}. The first term $ \mathbf{D}_{\mathrm{wFM}} \propto (\boldsymbol{\omega}_i - \boldsymbol{\omega}_j) \cdot (\mathbf{S}_i \times \mathbf{S}_j)$ couples the oxygen octahedra tilts $\boldsymbol{\omega}$ to the magnetic texture and stabilizes the magnetic moment in the $(111)$-plane perpendicular to the tilts rotation vectors. This DM contribution induces the weak ferromagnetic moment in BFO \cite{Ederer2005-sr}. The second term has the form $ \mathbf{D}_{\mathrm{SC}} \propto (\mathbf{u}_i \times \mathbf{e}_{ij}) \cdot (\mathbf{S}_i \times \mathbf{S}_j)$ and couples the polarization direction $\mathbf{u}_i$ to the magnetic moments and favors the stabilization of the magnetic moments perpendicular to the $(111)$ plane. By varying these energy terms, several types of spin cycloids have been predicted in BFO \cite{Xu2018-vc}.

To compare the energies of these different spin spirals, both DM contributions must be computed from DFT. $ \mathbf{D}_{\mathrm{wFM}}$ was first predicted to create the weak magnetic moment in BFO. Therefore it should lie along the $[111]$ direction \cite{Ederer2005-sr}. This DM contribution was obtained from DFT calculations (146~$\mu \mathrm{eV}$ \cite{Weingart2012-lb} up to 304~$\mu \mathrm{eV}$ \cite{Dixit2015-lf}) in relative good agreement with the experimental value (163~$\mu \mathrm{eV}$ \cite{Matsuda2012-bm}). Note that a full parametrization of BFO in the $R3c$ phase obtained from experimental measurements is given in Refs. \onlinecite{Park2014-le,Fishman2018}. They found $ \mathbf{D}_{\mathrm{SC}} = 110\,\mu \mathrm{eV}$ and $ \mathbf{D}_{\mathrm{wFM}} = 50\,\mu \mathrm{eV}$ which are significantly lower than previous measurements and calculations. To our surprise and to the best of our knowledge, $ \mathbf{D}_{\mathrm{SC}}$ has not yet been obtained from DFT calculations which does not allow to discuss the stability of the different spin spirals.

Here, we determine the origin and the strength of the spin current DM interaction in \BFO from DFT and show that in this type of multiferroics, the DM interaction originates from the asymmetric potential within the Fe cations rather than the structural distortions induced by the O anionic octahedra.
We calculate the energies of spin spirals $E(\mathbf{q})$ for different propagation directions $\mathbf{q}$ in the full pseudo-cubic BZ which shows an isotropic exchange interaction.
The spin-orbit coupling (SOC) contribution is fully quenched for spin spirals propagating along the polarization direction which only agrees with the spin current model \cite{Katsura2005,Rahmedov2012} and excludes the Levy and Fert model \cite{FertLevy1980}.
Finally, all magnetic interactions e.g. the magnetic exchange, the spin current model and the anisotropy lead to a degeneracy of spin cycloids within the $(111)$ plane which suggest that both type-I and type-II cycloid could coexist in bulk \textit{R3c} BFO. 

\begin{figure}
\centering
\includegraphics[scale=1]{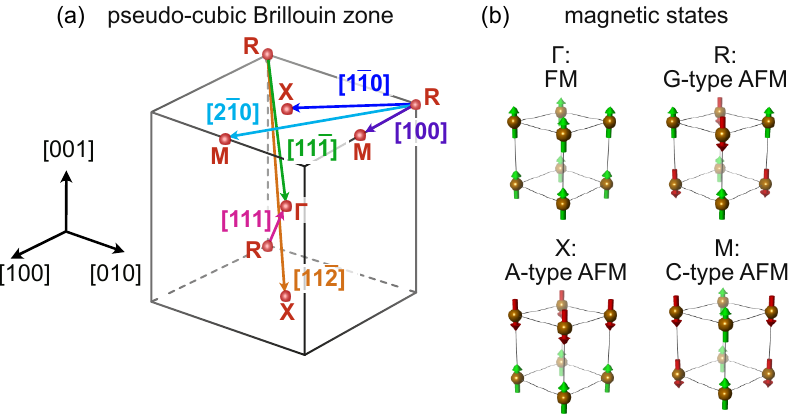}
\caption{(a) Pseudo-cubic Brillouin zone (BZ) with high symmetry points (red points and letters). Shown are the calculated propagation directions (colored arrows) of spin spirals. Note that due to representation, several R points are shown, the calculated directions, however, start from the same R points. (b) Collinear magnetic states (ferromagnetic -- FM, antiferromagnetic -- AFM) connected to each high symmetry point of the  BZ.}
\label{Figure: BFO BZ}
\end{figure}

\paragraph*{Methods} We have calculated the energy dispersions $E(\mathbf{q})$ of flat spin spiral states. These are the general solution of the Heisenberg model on a periodic lattice and can be characterized by the spin spiral vector $\mathbf{q}$. This vector determines the propagation direction of the spin spiral. A magnetic moment $\mathbf{S}_i$ at an atom position $\mathbf{r}_i$ is given by $\mathbf{S}_i = m \Big( \cos (\mathbf{q}\cdot \mathbf{r}_i),\sin (\mathbf{q}\cdot \mathbf{r}_i),0 \Big)$
where $m$ is the magnitude of the magnetic moment (see supplemental material \cite{Supplement}).
We choose $\mathbf{q}$ along several high symmetry directions of the rhombohedral BZ. For simplification, we present all data within the pseudo-cubic Brillouin zone, shown in Figure \ref{Figure: BFO BZ} (a). The different directions are drawn as colored arrows, starting from the R point. Every high symmetry point in the BZ is connected to a certain collinear state -- sketched in Fig.~\ref{Figure: BFO BZ} (b).

\begin{figure}
\centering
\includegraphics[scale=1]{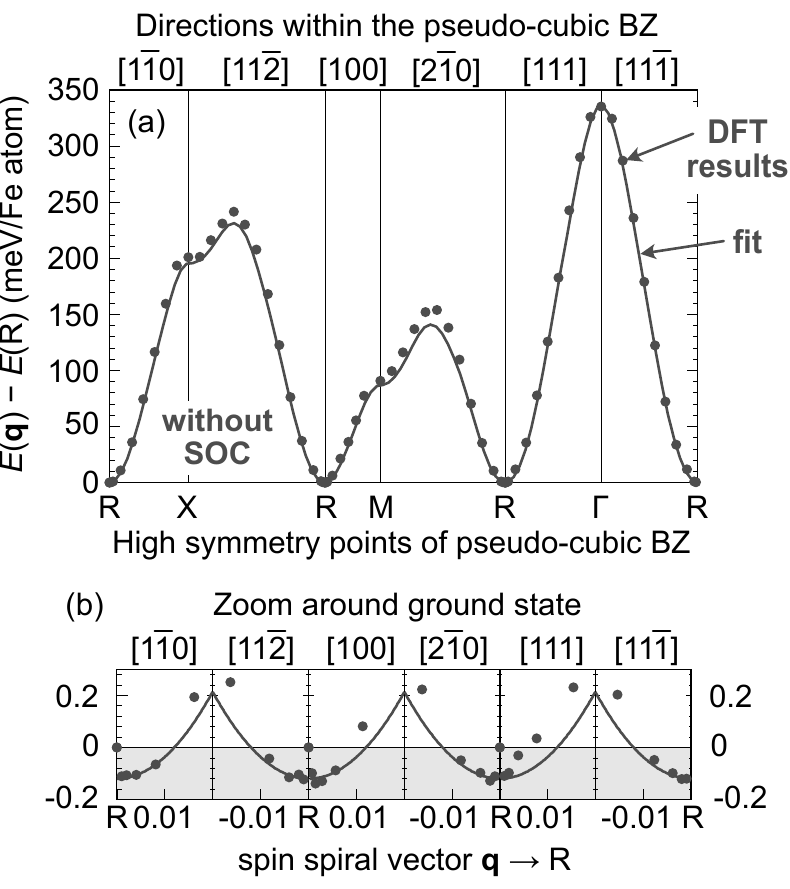}
\caption{Energy dispersion $E(\mathbf{q})$ without SOC of flat spin spirals along different directions within the pseudo-cubic Brillouin zone. Points show calculated energies from DFT, lines represent a fit to the Heisenberg exchange interaction including seven neighbors. (a) full energy dispersion. (b) Zoom around the ground state at $\mathbf{q}\rightarrow R$.}
\label{Figure: BFO exchange}
\end{figure}

\paragraph*{Results} The calculated energy dispersion (points) obtained without SOC is presented in Fig.~\ref{Figure: BFO exchange} (a) where the paths are $\mathrm{R}\rightarrow \mathrm{X}\rightarrow \mathrm{R}\rightarrow \mathrm{M}\rightarrow \mathrm{R}\rightarrow \mathrm{\Gamma}\rightarrow \mathrm{R}$ (cf. lower $x$ axis). On the upper $x$ axis, the respective directions according to Fig.~\ref{Figure: BFO BZ} can be seen. Without SOC, the dispersion shows an energy minimum at the R point (G-type AFM state) \cite{Rpoint} and the $\mathrm{\Gamma}$ point (the FM state) the highest energy. The energy dispersion is mapped onto an extended Heisenberg model $\mathcal{H}_\textrm{ex} = -\sum_{ij} J_{ij} (\mathbf{S}_i \cdot \mathbf{S}_j)$ to determine the strength of the Heisenberg exchange interaction parameters $J_{ij}$ [lines in panel (a) and (b)] beyond nearest neighbors (see supplemental material \cite{Supplement} for detailed information). The exchange between nearest neighbors is dominant $\sim -27\,\textrm{meV/Fe atom}$ capturing the large energy differences between the high symmetry points, but seven neighbors are necessary to describe the curvature around the R points (values can be found in Table~\ref{Table: values}). A closeup is shown in Fig.~\ref{Figure: BFO exchange} (b). The exchange interaction results in a good description of the DFT calculations. Note that the energy differences are very small and a numerical error on this energy scale is expected. Nevertheless, all the energies (data and fit) around the collinear state at the R points have the same curvature, e.g. the same effective exchange interaction \cite{Dupe2016}. Hence, the exchange in BFO is fully isotropic.

\begin{figure}
\centering
\includegraphics[scale=1]{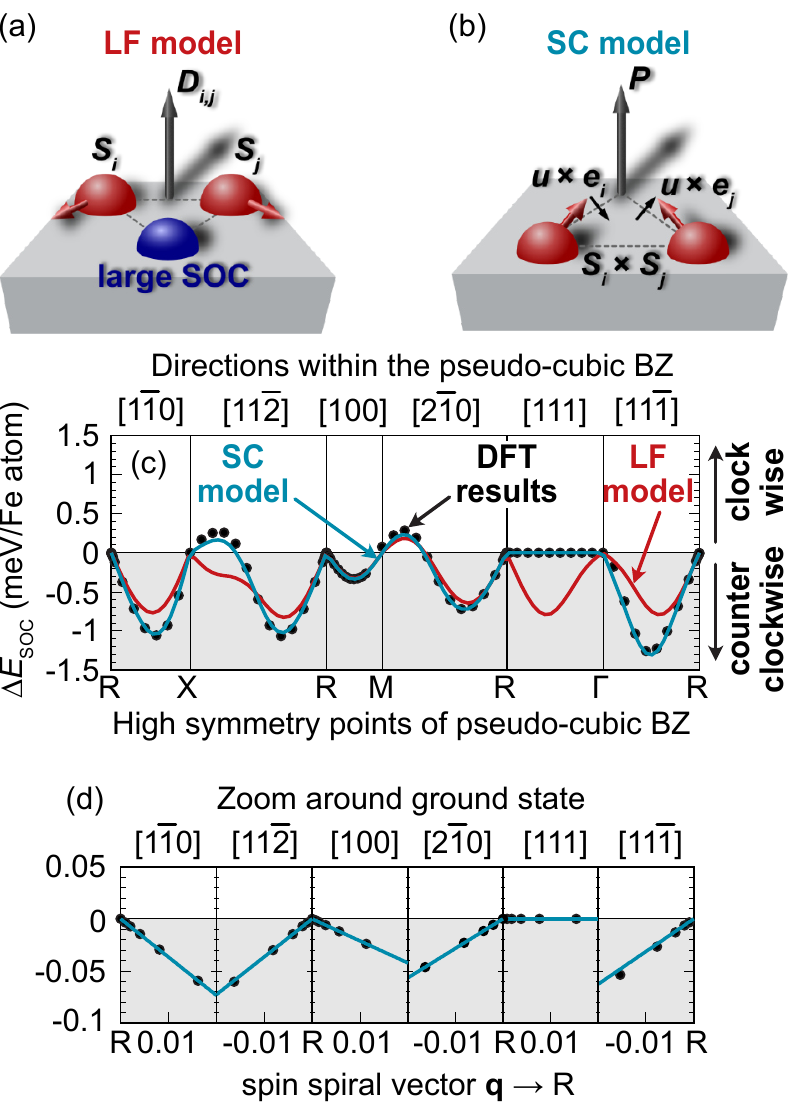}
\caption{Dzyaloshinskii-Moriya interaction (DM interaction) in BiFeO\sub{3}. (a) Model of Levy and Fert (LF model) to describe the Dzyaloshinskii-Moriya interaction \cite{FertLevy1980,LevyFert2016}. Two magnetic atoms $\mathbf{S}_i, \mathbf{S}_j$ are interacting via a heavy metal atom hosting large spin-orbit coupling (SOC). The triangular between the three atoms defines the DM vector $\mathbf{D}_{ij}$. (b) Converse spin-current (SC) model \cite{Katsura2005,Rahmedov2012}. In systems with a polarization $\mathbf{P}$, the spin-current vector $\mathbf{u}\times \mathbf{e}_{i,j}$ is perpendicular to the direction of polarization. (c) Energy contribution due to spin-orbit coupling $\Delta E_{\textrm{SOC}}$ to the energy dispersion of spin spirals along different directions within the pseudo-cubic Brillouin zone [Fig.~\ref{Figure: BFO BZ} (a)]. Points show calculated energies from density functional theory (DFT), the red line corresponds to the LF model, the blue line corresponds to the SC model. Positive (negative) values prefer clockwise (counter-clockwise) sense of rotation of non-collinear states. (d) Zoom around the ground state at $\mathbf{q}\rightarrow R$ where only the DFT data and the SC model is presented.}
\label{Figure: BFO DMI total}
\end{figure}

To determine the strength of the Dzyaloshinskii-Moriya (DM) interaction, we apply spin-orbit coupling (SOC) in first order perturbation \cite{Heide2009} to every calculated point of Figure~\ref{Figure: BFO exchange} \cite{Supplement}. The resulting energy contribution due to SOC, $\Delta E_{\textrm{SOC}}$ is illustrated in Figure~\ref{Figure: BFO DMI total}. While in some directions with higher symmetry [panel (c)], such as $[1\overline{1}0], [100] \, \textrm{and} \, [11\overline{1}]$ the DFT results (black points) show a simple sine behavior, other directions ($[11\overline{2}]\, \textrm{and}\, [2\overline{1}0]$) exhibit a more complex trend. In the [111] direction -- the direction of spontaneous polarization in \BFO -- however the energy of SOC is completely quenched.

To quantify the strength of the DM interaction, we apply the model of Levy and Fert (LF model) \cite{FertLevy1980} which is sketched in panel (a) of Fig.~\ref{Figure: BFO DMI total}. The LF model is typically explained as a superexchange mechanism between two magnetic atoms $\mathbf{S}_i, \mathbf{S}_j$ via a third non-magnetic atom that holds a large spin-orbit coupling. Due to the direction dependent scattering of the electrons, non-collinear magnetic structures with a specific sense of rotation (clockwise or counterclockwise) are preferred. Using the symmetry of the DM vector $\mathbf{D}_{ij} \propto \frac{\mathbf{R}_i \times \mathbf{R}_j}{\vert \mathbf{R}_i \times \mathbf{R}_j  \vert}$ where $\mathbf{R}_{i,j}$ are the position vectors of magnetic moments $\mathbf{S}_{i,j}$ with respect to the atom of large SOC, it is possible to fit every direction in panel (c) separately and to determine the DM interaction (cf. \cite{Supplement}). This is the reason, why in other publications, the magnitude for the DM vector can be obtained from first principles. However, since our calculations include different high symmetry directions including the [111] direction, a combined mapping of the DM vector to our data is not reproducing the DFT calculations [see red line in Fig.~\ref{Figure: BFO DMI total} (c)]. Hence, the model of Levy and Fert cannot be used to describe the DM interaction in BFO. Note that the DM interaction $\mathbf{D}_\textrm{wFM}$ which couples the oxygen octahedra tilts to the magnetic structure and explains the weak FM canting coincides with the LF model. It would give rise to an energy contribution of SOC in the [111] direction. However, our tests reveal completely vanishing energies in the direction of polarization.

In the literature, another term is shown to describe the non-collinear canting in \BFO, the so called (converse) spin-current model \cite{Katsura2005,Rahmedov2012,Xu2018,Burns2019}. This magneto-electric effect is based on the spin current acting on non-collinear magnets and leads to the polarization $ \mathbf{P} \propto \mathbf{e}_{ij} \times \left ( \mathbf{S}_i \times \mathbf{S}_j \right )$. Following the convention of Ref.~\onlinecite{Rahmedov2012}, the SC DM interaction is described via
\begin{equation}
    \mathcal{H}_\textrm{SC}^\textrm{DMI} = - \sum_{ij} C_{ij} \left ( \mathbf{u} \times \mathbf{e}_{ij} \right ) \cdot \left ( \mathbf{S}_i \times \mathbf{S}_j \right ) \label{Equation: Spin Current model}
\end{equation} 
where $\mathbf{u}$ is the unit vector in the direction of polarization, $\mathbf{e}_{ij}$ the unit vector connecting magnetic moments $\mathbf{S}_{i,j}$ at sites $i,j$. The parameter $C_{ij}$ describes the strength of the DM interaction. The difference between the DM vector $\mathbf{D}_{ij}$ and the SC vector $C_{ij} \left ( \mathbf{u} \times \mathbf{e}_{ij} \right )$ lies within the symmetry. Depending on the symmetry of the system however, both vectors can coincide, e.g. in two-dimensional interfaces. In \BFO both vectors differ -- most noticeably in the [111] direction -- and hence the two models lead to different results modeling the DM interaction. 

Mapping Eq.~\eqref{Equation: Spin Current model} to the energy contribution $\Delta E_\textrm{SOC}$ gives rise to the blue lines in Fig.~\ref{Figure: BFO DMI total} (c) and (d). The SC model describes the DFT results almost perfectly. Here, the fit contains three neighbors, where the contributions of the second and the third neighbors are a minor correction to the first neighbor (cf. Tab.~\ref{Table: values}). Note that even an effective nearest neighbor approximation for $C_{ij}$, cf. $C_\textrm{eff}$ shows a qualitatively very good agreement with the DFT calculations (cf. supplemental material \cite{Supplement}).

In panel (d) of Figure~\ref{Figure: BFO DMI total}, we see the energy differences between the different pseudo-cubic propagation directions of spin spirals around the ground state. Both data and SC model exhibit the steepest slope in the $[1\overline{1}0]\, \textrm{and}\, [11\overline{2}]$ directions meaning the DM interaction for these two directions is the largest. Furthermore, for both directions, points and lines show complete energy degeneracy. $[1\overline{1}0]\, \textrm{and}\, [11\overline{2}]$ referring typically to the type-I and type-II cycloids, respectively, lie within the (111) plane. Our tests reveal that any direction within the (111) plane is equally preferred by the spin-current DM interaction. That means, including exchange and spin-current DM interaction restricts the possibility of spin cycloid propagation directions in \BFO bulk to any direction of the two-dimensional (111) plane. This is in accordance with Ref.~\onlinecite{Fishman2018}, where it is also stated that $\mathbf{q}$ can point along any direction in the hexagonal (111) plane.

\begin{figure}
\centering
\includegraphics[scale=1]{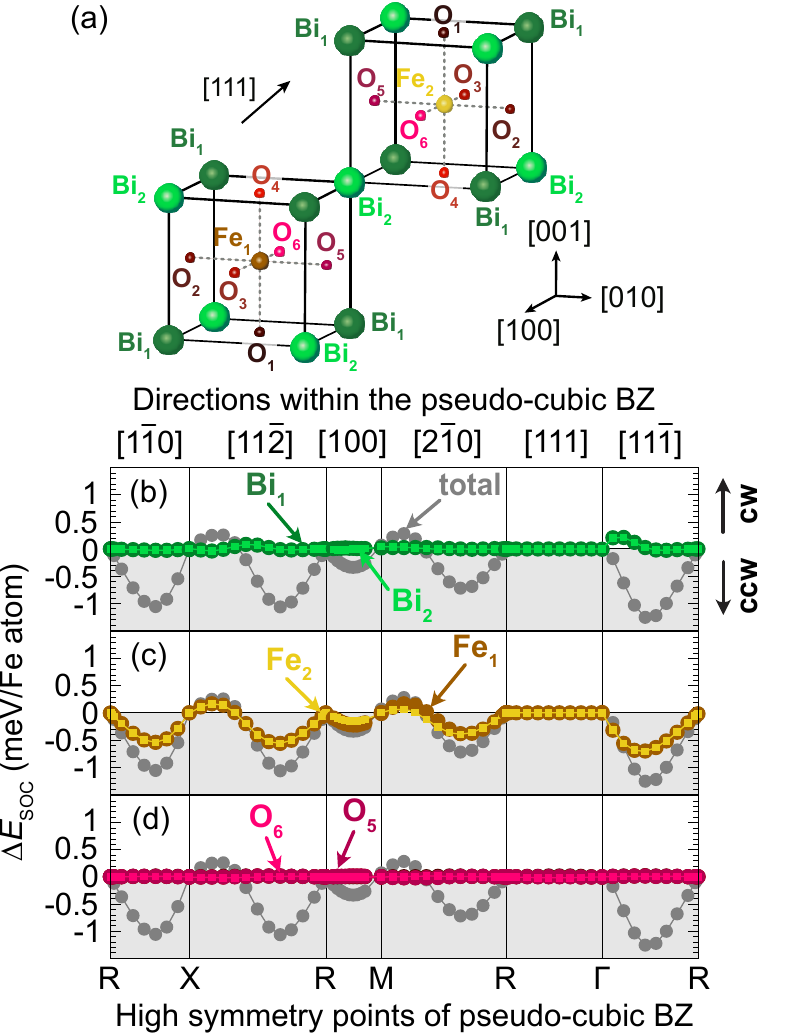}
\caption{Element resolved energy contribution due to SOC $\Delta E_\textrm{SOC}$ to the energy dispersion. (a) Sketch of \BFO structure in cubic approximation. (b,c,d) $\Delta E_\textrm{SOC}$ for both Bi, both Fe and six O atoms in the unit cell, respectively. The total energy contribution is shown in gray.}
\label{Figure: BFO DMI element}
\end{figure}

For a detailed insight into the origin of the DM interaction in BFO, we calculated the element resolved energy contribution due to SOC (Fig.~\ref{Figure: BFO DMI element}). Here, the gray points show the total energy contribution from Fig.~\ref{Figure: BFO DMI total} (c) and the colored points represent the elements of \BFO (lines serve as guide to the eye). In panels (b), (c), (d) the energy contributions due to Bi, Fe and O, respectively are presented. Since in the rhombohedral unit cell, two Bi, two Fe and six O atoms are used, a sketch of \BFO in cubic representation is shown in (a) with the same color code for each element as in the graphs. Even though Bi is the element of large SOC and the O atoms break the inversion symmetry, their contribution to the total DM interaction in BFO is negligible [panels (b) and (d)]. This emphasizes that the model of Levy and Fert relying on an heavy metallic element with large SOC and structural asymmetry does not provide a good description for BFO. The total DM interaction -- as a sum of all contributions -- is represented by the whole contribution of both Fe atoms. This is similar to the Rashba DM interaction observed in on 3\textit{d} unsupported monolayers (UMLs) under an external electric field \cite{Desplat2021} and in Graphene/Co(0001) due to potential gradient between graphene and Co \cite{Yang2018-qm}. It creates an internal asymmetry of the potential leading to a non-vanishing DM interaction even though the structure itself does not hold an asymmetry. In BFO, the internal electric field creates the same asymmetry within the potential of the Fe atoms and hence the DM interaction is driven by the spin current of the non-collinear magnetic structure. 

As mentioned above, the exchange is isotropic within the full 3D BZ, whereas the spin current driven DM interaction narrows the spin cycloid directions down to the (111) plane. However, experiments on bulk \textit{R3c} BFO only observe type-I cycloids, propagating in $[1\overline{1}0]$ and the two equivalent directions $[\overline{1}01]$ and $[01\overline{1}]$. Therefore, we determine the uniaxial anisotropy energy to check whether this interaction can pin the cycloid's propagation down to these three directions. In Ref.~\onlinecite{Xu2021} it has been shown that the small easy (111) plane anisotropy in \textit{R3c} BFO is a result of both strong out-of-plane anisotropy (in [111] direction) driven by the Bi-Fe ferroelectric displacement in competition with a strong easy plane anisotropy (perpendicular to the [111] direction) stemming from the oxygen octahedra tilts.
Here, we calculate the anisotropy energy using a $360^\circ$ rotation within the (111) plane. The magnetocrystalline anisotropy energy (MAE) is not changing for any direction in this plane \cite{Supplement}. In bulk BFO, the MAE does not prefer a certain magnetization direction within the (111) plane.

\begin{table}
\centering
\caption{Magnetic interactions in BiFeO\sub{3} mapping an atomistic spin model to the results of DFT calculations. All values of the $i$-th neighbour exchange $J_i$, Dzyaloshinskii-Moriya constant of the spin-current model $C_{ij}$ and uniaxial magnetocrystalline anisotropy $K$ are given in meV/Fe atom. $J > 0$ ($J<0$) represents FM (AFM) order, $C > 0 $ ($C<0$)  counterclockwise (clockwise) rotation. $K > 0$ prefers an easy plane perpendicular to the [111] direction.}\label{Table: values}
\begin{tabular}{cc} 
Parameter & value (meV/Fe atom) \\ \hline
$J_1$   &   $-26.998$ \\
        $J_2$   &   $-2.005$  \\
        $J_3$   &   $-0.179$  \\
        $J_4$   &   $+0.717$ \\ 
        $J_5$   &   $-0.195$  \\
        $J_6$   &   $+0.157$  \\
        $J_7$   &   $-0.410$  \\ \hline
        $C_1$   & $+0.369$ \\
        $C_2$   & $-0.00360$ \\
        $C_3$   & $-0.0156$ \\ \hline
        $C_\textrm{eff}$ & $+0.301$ \\ \hline
        $K$     &   $+0.054$ \\
\end{tabular}
\end{table}

\paragraph*{Conclusion} Our work demonstrates based on density functional theory that the Dzyaloshinskii-Moriya interaction in \textit{R3c} bulk \BFO is governed by the spin-current model stemming from the non-collinear antiferromagnetic structure. The off-centered displacement of Fe and Bi induces an asymmetric shape of the internal potential in the Fe atoms and consequently, the Fe atoms carry almost the whole contribution to the DM interaction. This effect is not restricted to BFO in particular and should occur in other multiferroic materials.
In the case of BFO, by including exchange, DM interaction and anisotropy, we show that any spin cycloid propagation direction in the (111) plane are energetically degenerate.
This could explain an effect of continuously rotating cycloids in BFO that has recently been observed \cite{Finco2022} and explained as a surface effect. Based on our results, this effect should also be measurable in bulk.

\begin{acknowledgments}
S.M., B.D. thank Dr. Manuel Bibes and Dr. Eric Bousquet for fruitful discussions and careful reading of the manuscript.
This work is supported by the National Natural Science Foundation of China under Grant No. 12074277, the startup fund from Soochow University and the support from Priority Academic Program Development (PAPD) of Jiangsu Higher Education Institutions. S.M., M.J.V., B.D., and L.B. acknowledge the DARPA Grant No. HR0011727183-D18AP00010 (TEE Program) and the European Union’s Horizon 2020 research and innovation programme under Grant Agreements No. 964931 (TSAR). L.B. also thanks ARO Grant No. W911NF-21-1-0113 and ARO Grant No. W911NF-21-2-0162 (ETHOS). Computing time was provided by ARCHER and ARCHER2 based in the United Kingdom at National Supercomputing Service with support from the PRACE aisbl, and from the Consortium d’Equipements de Calcul Intensif (FRS-FNRS Belgium GA 2.5020.11).
\end{acknowledgments}


\begin{thebibliography}{39}
\bibitem [{Rpo()}]{Rpoint}%
	\BibitemOpen
	\href@noop {} {}\bibinfo {note} {Note that in the data without SOC, we see a
		tiny energy difference between the perfectly collinear state (R points) and
		the next states. This energy difference stems from the rotation of
		magnetization as it is implemented in the FLAPW method.}\BibitemShut {Stop}%
\end{thebibliography}
\end{document}